\documentclass[prl,twocolumn,noshowpacs,nobalancelastpage,tightenlines,floatfix]{revtex4}
\usepackage{amssymb}
\usepackage{graphicx}

\begin{document}

\title{Observation of the Vacuum-Rabi Spectrum for One Trapped Atom}
\author{A. Boca, R. Miller, K.~M. Birnbaum, A.~D. Boozer, J. McKeever, and
H.~J. Kimble}
\affiliation{Norman Bridge Laboratory of Physics 12-33\\
California Institute of Technology, Pasadena, CA 91125}
\date{\today}

\begin{abstract}
The transmission spectrum for one atom strongly coupled to the field of a
high finesse optical resonator is observed to exhibit a clearly resolved
vacuum-Rabi splitting characteristic of the normal modes in the eigenvalue
spectrum of the atom-cavity system. A new Raman scheme for cooling atomic
motion along the cavity axis enables a complete spectrum to be recorded for
an individual atom trapped within the cavity mode, in contrast to all
previous measurements in cavity QED that have required averaging over many
atoms.
\end{abstract}

\maketitle

A cornerstone of optical physics is the interaction of a single two-level
atom with the electromagnetic field of a high quality resonator. Of
particular importance is the regime of strong coupling, for which the
frequency scale $g$ associated with reversible evolution for the atom-cavity
system exceeds the rates $(\gamma ,\kappa )$ for irreversible decay of atom
and cavity field, respectively \cite{kimble98}. In the domain of strong
coupling, a photon emitted by the atom into the cavity mode is likely to be
repeatedly absorbed and re-emitted at the single-quantum Rabi frequency $2g$
before being irreversibly lost into the environment. This oscillatory
exchange of excitation between atom and cavity field results from a normal
mode splitting in the eigenvalue spectrum of the atom-cavity system \cite%
{jaynes63} which is manifest in emission \cite{sanchez83} and absorption
\cite{agarwal84} spectra, and has been dubbed the vacuum-Rabi splitting \cite%
{sanchez83}.

Strong coupling in cavity QED as evidenced by the vacuum-Rabi splitting
provides enabling capabilities for quantum information science, including
for the implementation of scalable quantum computation \cite%
{pellizzari95,duan04}, for the realization of distributed quantum networks
\cite{cirac97,briegel00}, and more generally, for the study of open quantum
systems \cite{mabuchi02}. Against this backdrop, experiments in cavity QED
have made great strides over the past two decades to achieve strong coupling
\cite{berman94}. The vacuum-Rabi splitting for single intracavity atoms has
been observed with atomic beams in both the optical \cite%
{thompson92,childs96,thompson98}\ and microwave regimes
\cite{brune96}. The combination of laser cooled atoms and large
coherent coupling has enabled single atomic trajectories to be
monitored in real time with high signal-to-noise ratio, so that the
vacuum-Rabi spectrum could be obtained from atomic transit signals
produced by single atoms \cite{hood98}. A significant advance has
been the trapping of individual atoms in an optical cavity in a
regime of strong coupling \cite{ye99,mckeever03}, with the
vacuum-Rabi splitting first evidenced for single trapped atoms in
Ref. \cite{ye99} and the entire transmission spectra recorded in
Ref. \cite{maunz04}.

Without exception these prior single atom experiments related to
the vacuum-Rabi splitting in cavity QED \cite%
{thompson92,brune96,childs96,thompson98,hood98,ye99,mckeever03,maunz04} have
required averaging over trials with many atoms to obtain quantitative
spectral information, even if individual trials involved only single atoms
(e.g., $10^{5}$ atoms were required to obtain a spectrum in Ref. \cite%
{brune96} and $>10^{3}$ atoms were needed in Ref. \cite{maunz04}).
By contrast, the implementation of complex algorithms in quantum
information science requires the capability for repeated
manipulation and measurement of an individual quantum system, as has
been spectacularly demonstrated with trapped ions
\cite{riebe04,barrett04} and recently with Cooper pair boxes
\cite{wallraff04}.

\begin{figure}[tb]
\includegraphics[width=6.45cm]{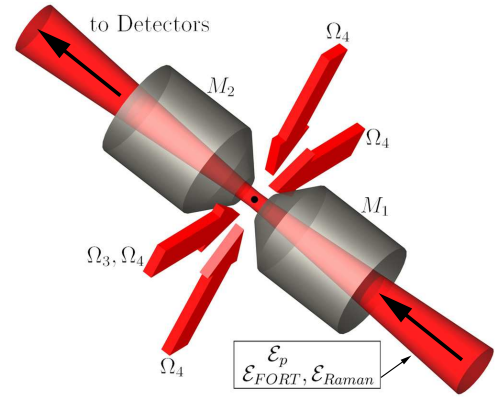}
\caption{A single atom is trapped inside an optical cavity in the regime of
strong coupling by way of an intracavity FORT driven by the field $\mathcal{%
E }_{FORT}$. The transmission spectrum $T_{1}(\protect\omega _{p})$
for the atom-cavity system is obtained by varying the frequency
$\protect\omega _{p}$ of the probe beam $\mathcal{E}_{p}$ and
recording the output with single-photon detectors. Cooling of the
radial atomic motion is accomplished with the transverse fields
$\Omega _{4}$, while axial cooling results from Raman transitions
driven by the fields $\mathcal{E} _{FORT}$, $\mathcal{E}_{Raman}$.
An additional transverse field $\Omega_3$ acts as a repumper during
probe intervals.} \label{fig1}
\end{figure}

With this goal in mind, in this Letter we report measurements of
the spectral response of single atoms that are trapped and
strongly coupled to the field of a high finesse optical resonator.
By alternating intervals of probe measurement and of atomic
cooling, we record a complete probe spectrum for one and the same
atom. The vacuum-Rabi splitting is thereby measured in a
quantitative fashion for each atom by way of a protocol that
represents a first step towards more complex tasks in quantum
information science. An essential component of our protocol is a
new Raman scheme for cooling atomic motion along the cavity axis,
that leads to inferred atomic localization $\Delta z_{axial}\simeq
33~\mathrm{nm}$, $\Delta \rho _{transverse}\simeq 5.5~\mathrm{\mu
m}$.

A simple schematic of our experiment is given in Fig.~\ref{fig1},
with technical details provided in Ref. \cite{mckeever-thesis}.
After release from a magneto-optical trap (MOT) located several mm
above the Fabry-Perot cavity formed by mirrors $(M_{1},M_{2})$,
single Cesium atoms are cooled and loaded into an intracavity
far-off-resonance trap (FORT) and are thereby
strongly coupled to a single mode of the cavity. Our experiment employs the $%
6S_{1/2},F=4\rightarrow 6P_{3/2},F^{\prime }=5^{\prime }$ transition of the $%
D2$ line in Cs at $\lambda _{A}=852.4~\mathrm{nm}$, for which the maximum
single-photon Rabi frequency $2g_{0}/2\pi =68~\mathrm{MHz}$ for $%
(F=4,m_{F}=\pm 4)\rightarrow (F^{\prime }=5^{\prime},m_{F}^{\prime
}=\pm 5)$. The transverse decay rate for the $6P_{3/2}$ atomic
excited states is $\gamma /2\pi =2.6~\mathrm{MHz}$, while the
cavity field decays at rate $\kappa /2\pi =4.1~\mathrm{MHz}$
\cite{hood01}. Our system is in the strong coupling regime of
cavity QED $g_{0}\gg (\gamma ,\kappa )$ \cite{kimble98}, with
critical photon and atom numbers $n_{0}\equiv \gamma
^{2}/(2g_{0}^{2})\approx 0.0029$ and $N_{0}\equiv 2\kappa \gamma
/g_{0}^{2}\approx 0.018$.

The intracavity FORT is driven by a linearly polarized input field $\mathcal{%
E}_{FORT}$ at $\lambda _{F}=935.6~\mathrm{nm}$ \cite{birefringence},
resulting in nearly equal ac-Stark shifts for all Zeeman states in the $%
6S_{1/2},F=3,4$ manifold \cite{corwin99}. At an antinode of the field, the
peak value of the trapping potential for these states is $U_{0}/h=-39~%
\mathrm{MHz}$ for all our measurements. Zeeman states of the $%
6P_{3/2},F^{\prime }=5^{\prime}$ manifold likewise experience a
trapping potential, albeit with a weak dependence on
$m^{\prime}_F$ \cite{mckeever03}.
The cavity length is independently stabilized to length $l_{0}=42.2~\mathrm{%
\mu m}$ such that a $TEM_{00}$ mode at $\lambda _{C_{1}}$ is resonant with
the free-space atomic transition at $\lambda _{A}$ and another $TEM_{00}$
mode at $\lambda _{C_{2}}$ is resonant at $\lambda _{F}$. The cavity waists
are $w_{C_{1,2}}=\{23.4~\mathrm{\mu m},~24.5~\mathrm{\mu m}\}$ at $\lambda
_{C_{1,2}}=\{852.4~\mathrm{nm},~935.6~\mathrm{nm}\}$.

As illustrated in Fig.~\ref{fig1}, we record the transmission spectrum $%
T_{1}(\omega _{p})$ for a weak external probe $\mathcal{E}_{p}$ of variable
frequency $\omega _{p}$ incident upon the cavity containing one strongly
coupled atom \cite{T-define}. Our protocol consists of an alternating
sequence of probe and cooling intervals. The probe beam is linearly
polarized \cite{probepol} and is matched to the $TEM_{00}$ mode around $\lambda _{C_{1}}$. $%
\mathcal{E}_{p}$ illuminates the cavity for $\Delta t_{probe}=$ $100~\mathrm{%
\mu s}$, and the transmitted light is detected by two avalanche
photodiodes for photon counting \cite{efficiency}. During this
interval a repumping beam $\Omega _{3}$, transverse to the cavity
axis, of frequency resonant with $6S_{1/2},F=3\rightarrow
6P_{3/2},F^{\prime }=4^{\prime }$, also illuminates the atom. In
successive probe intervals, the frequency $\omega _{p}$ of the probe
beam is swept with an approximately linear ramp from well below to
far above the common atom-cavity resonance at $\omega _{A}\simeq
\omega _{C_{1}}$. The frequency sweep for the probe beam is repeated
eight times in $\Delta t_{tot}=1.2~\mathrm{s}$, and then a new
loading cycle is initiated.

Following each probe interval, we apply light to cool both the
radial and axial motion for $\Delta t_{cool}=$ $2.9~\mathrm{ms}$.
Radial cooling is achieved by the $\Omega _{4}$\ beams consisting
of two pairs of counter-propagating fields in a $\sigma _{\pm }$
configuration perpendicular to the cavity axis, as shown in
Fig.~\ref{fig1}. The $\Omega _{4}$\ beams
are detuned $\Delta _{4}\simeq 10~\mathrm{MHz}$ to the \textit{blue} of the $%
4\rightarrow 4^{\prime }$ transition to provide blue Sisyphus cooling \cite%
{boiron96} for atomic motion transverse to the cavity axis.

\begin{figure}[tb]
\includegraphics[width=8.6cm]{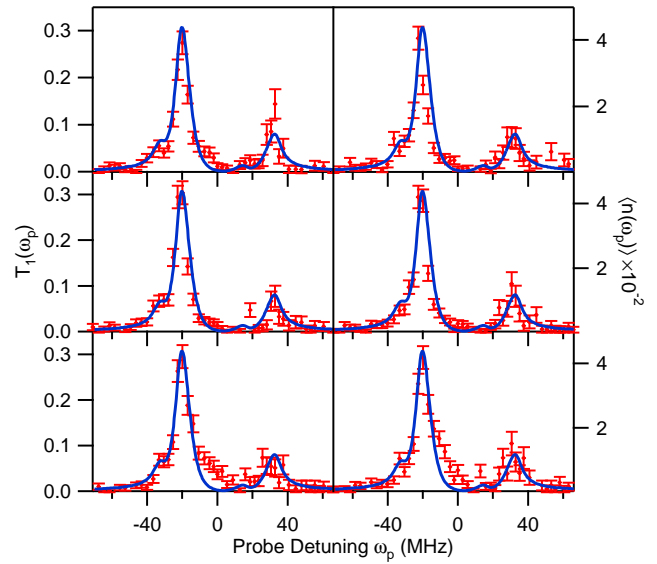}
\caption{Transmission spectrum $T_{1}(\protect\omega _{p})$ for six randomly
drawn atoms \protect\cite{T-define}. In each case, $T_{1}(\protect\omega %
_{p})$ is acquired for one-and-the-same atom, with the two peaks of
the vacuum-Rabi spectum clearly evident. The error bars reflect the
statistical uncertainties in the number of photocounts. The full
curve is from the steady-state solution to the master equation as
discussed in the text.} \label{fig2}
\end{figure}

To cool the axial motion for single trapped atoms, we have developed a new
scheme that employs $\mathcal{E}_{FORT}$ and an auxiliary field $\mathcal{E}%
_{Raman}$ that is frequency offset by $\Delta _{Raman}=\Delta _{HF}+\delta $
and phase locked to $\mathcal{E}_{FORT}$. Here, $\Delta _{HF}=9.192632~%
\mathrm{GHz}$ is the hyperfine splitting between the $6S_{1/2},F=3,4$ levels
and $\delta $ is a detuning. $\mathcal{E}_{FORT},~\mathcal{E}%
_{Raman} $ generate intracavity fields with Rabi frequencies
$\Omega _{FORT},~\Omega _{Raman}$ that drive Raman transitions
between the $F=3,4$ levels with effective Rabi
frequency $\Omega _{E}\approx\Omega _{FORT}\Omega _{Raman}/(2\Delta)$%
, where in a simple theory the detuning $\Delta =\omega _{A}-\omega _{FORT}$%
. In our experiments, $\Omega _{E}\sim 200~\mathrm{kHz}$. By tuning $\delta $
near the $\Delta n=-2$ motional sideband (i.e., $-2\nu _{0}\sim\delta=-1.0~%
\mathrm{MHz}$ with $\nu _{0}$ as the vibrational frequency for harmonic
motion in the axial direction around an antinode of the FORT), we implement
sideband cooling via the $F=3\rightarrow 4$ transition, with repumping
provided by the radial $\Omega _{4}$\ beams. Note that the Raman process
also acts as a repumper for population pumped to the $F=3$ level by the $%
\Omega_4$ beams. Each cooling interval is initiated by turning on
the fields $\Omega _{4},$ $\mathcal{E} _{Raman}$\ during $\Delta
t_{cool}$ and is terminated by gating these fields off before the
next probe interval $\Delta t_{probe}$.

Figure~\ref{fig2} displays the normalized transmission spectra
$T_{1}$ \cite{T-define} for individual atoms acquired by way of our
protocol of alternating probe and cooling intervals. Clearly evident
in each trace is a two-peaked structure that represents the
vacuum-Rabi splitting observed on an atom-by-atom basis. For
comparison, also shown in the figure is the predicted transmission
spectrum obtained from the steady-state solution to the master
equation for one atom strongly coupled to the cavity, as discussed
below. The quantitative correspondence between theory and experiment
is evidently quite reasonable for each atom \cite{asymmetry}.

To obtain the data shown in Fig.~\ref{fig2}, $N_{load}=61$ atoms
were loaded into the FORT in $500$ attempts, with the probability
that a given successful attempt involved $2$ or more atoms
estimated to be $P_{load}(N\geq 2)\lesssim 0.06$. Of
these $61$ atoms, $N_{survive}=28$ atoms remained trapped for the entire duration $%
\Delta t_{tot}$ of the measurement of $T_{1}(\omega _{p})$. The six spectra
shown in Fig.~\ref{fig2} were selected by a random drawing from this set of $%
N_{survive}$ atoms. Our sole selection criterion for presence of
an atom makes no consideration of the spectral structure of
$T_{1}(\omega _{p})$ except that there should be large absorption
on line center, $T_{1}(\omega _{p}=\omega _{C_{1}})\leq
T_{thresh}\approx 0.2$, with $T_{0}(\omega _{p}=\omega
_{C_{1}})\equiv1$ for the empty cavity \cite{selection}. Note that
the lifetime for an atom trapped in the FORT\ in the absence of
the cooling and probing light is $\tau
_{0}\simeq 3~\mathrm{s}$, which leads to a survival probability $%
p(\Delta t_{tot})\simeq 0.7$.

\begin{figure}[tb]
\includegraphics[width=8.6cm]{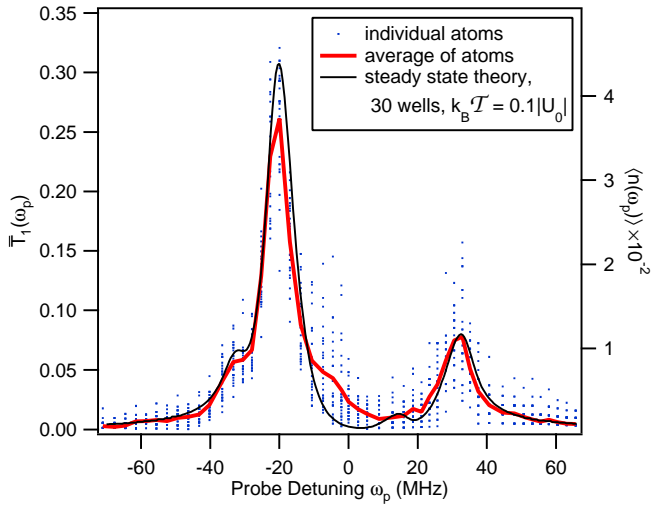}
\caption{Transmission spectrum $\bar{T}_{1}(\protect\omega _{p})$ resulting
from averaging individual spectra for $N=28$ atoms as in Fig.~\protect\ref%
{fig1} \protect\cite{T-define}. The points are a scatter plot from
the data for $T_{1}(\protect\omega _{p})$ for all $28$ atoms. The
thin trace is from the steady-state solution to the master equation,
and is identical to that in Fig.~\protect\ref{fig2}. The only free
parameters in the theory are the temperature and the range of FORT
antinodes; the vertical scale is absolute.} \label{fig3}
\end{figure}

In Fig.~\ref{fig3} we collect the results for $T_{1}(\omega _{p})$
for all $N_{survive}=28$ atoms, and display the average transmission
spectrum $\bar{T }_{1}(\omega _{p})$, as well a scatter plot from
the individual spectra. This comparison demonstrates that the
vacuum-Rabi spectrum observed for any particular atom represents
with reasonable fidelity the spectrum that would be obtained from
averaging over many atoms, albeit with fluctuations due to Poisson
counting and optical pumping effects over the finite duration of the
probe. Note that the total acquisition time associated with the
probe beam for the spectrum of any one atom is only $40~
\mathrm{ms}$, which can be improved as we optimize the cooling
protocol to give better statistics for each atom.

We have also acquired transmission spectra $T_{1}(\omega _{p})$
under operating conditions beyond those displayed in
Figs.~\ref{fig2},~\ref{fig3}, including drive intensities
$|\mathcal{E}_{p}|^{2}$ varied by factors of $2$, $\frac{1}{2}$, and
$\frac{1}{4}$, and atom-cavity detunings $\Delta
_{AC}=\omega_A-\omega_{C_1}=\pm 13~ \mathrm{MHz}$. In all cases, we
observe a distinctive vacuum-Rabi splitting on an atom-by-atom
basis, and will discuss the quantitative comparison of these results
with theory elsewhere \cite{birnbaum04}.

The full curves in Figs.~\ref{fig2},~\ref{fig3} are obtained from
the steady state solution of the master equation including all
transitions $(F=4,m_{F}) \leftrightarrow (F^{\prime
}=5^{\prime},m_{F}^{\prime})$ with their respective coupling
coefficients $g_{0}^{(m_{F},m_{F}^{\prime })}$, as well as the two
nearly degenerate modes of our cavity
\cite{birefringence,probepol}. For the comparison of theory and
experiment, the parameters $(g_{0}^{(m_{F},m_{F}^{ \prime
})},\gamma ,\kappa ,\omega_{C_1},\omega
_{p},|\mathcal{E}_{p}|^{2},U_0)$ are known in absolute terms
without adjustment. However, we have no \textit{a priori}
knowledge of the particular FORT well into which the atom is
loaded along the cavity standing wave, nor of the energy of the
atom. The FORT shifts and coherent coupling rate are both
functions of atomic position $\mathbf{r}$, with
$U(\mathbf{r})=U_{0}\sin ^{2}(k_{C_2}z)\exp (-2\rho
^{2}/w_{C_2}^{2})$ and $g^{(m_{F},m_{F}^{\prime })}(\mathbf{r}%
)=g_{0}^{(m_{F},m_{F}^{ \prime })}\psi (\mathbf{r})$, where $%
g_{0}^{(m_{F},m_{F}^{\prime })}=g_{0}G_{m_{F},m_{F}^{\prime }}$ with $%
G_{i,f} $ related to the Clebsch-Gordan coefficient for the particular $%
m_{F}\leftrightarrow m_{F}^{\prime }$ transition. $\psi
(\mathbf{r})=\sin (k_{C_1}z)\exp (-\rho ^{2}/w_{C_1}^{2})$, where
$\rho $ is the transverse distance from the cavity axis $z$, and
$k_{C_{1,2}}=2\pi /\lambda _{C_{1,2}}$.

As discussed in connection with Fig.~\ref{fig4} below, for the
theoretical curves shown in Figs.~\ref{fig2},~\ref{fig3}, we have
chosen only the $30$ out of $90$ total FORT wells for which $|\psi
(\mathbf{r}_{FORT})|\geq 0.87$,
where $\mathbf{r}_{FORT}$ specifies a maximum for the FORT intensity (i.e., $%
U(\mathbf{r}_{FORT})=U_{0}$). Furthermore, for these wells we have averaged $%
T_{1}(\omega _{p})$ over a Gaussian distribution in position $\mathbf{r}$
consistent with a temperature $k_{B}\mathcal{T}=0.1U_{0}$ ($\sim 200~%
\mathrm{\mu K}$). Since all parameters are known except for those
that characterize atomic motion, the good agreement between theory
and experiment \cite{T-disparity} allows us to infer that our
cooling protocol together with the selection criterion
$T_{thresh}=0.2$ results in individual atoms that are strongly
coupled
in one of the \textquotedblleft best\textquotedblright\ FORT wells (i.e., $%
|\psi (\mathbf{r}_{FORT})|\gtrsim 0.87$) with \textquotedblleft
temperature\textquotedblright\ $\sim 200~\mathrm{\mu K}~
$\cite{nonthermal}.

\begin{figure}[tb]
\includegraphics[width=8.6cm]{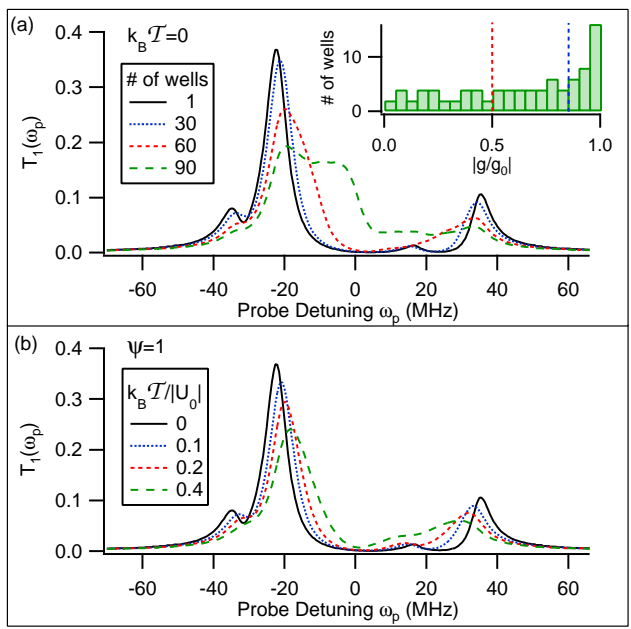}
\caption{Theoretical plots for $T_{1}(\protect\omega _{p})$ from the
steady-state solution to the master equation. (a) For zero
temperature, $T_{1}(\protect\omega _{p})$ is calculated from an
average over various FORT antinodes along the cavity axis, with the
inset showing the
associated distribution of values for $|g(\mathbf{r}%
_{FORT})|/g_{0}$. (b) For an optimum FORT well (i.e., $|g(%
\mathbf{r}_{FORT})|/g_{0}=1$ ), $T_{1}(\protect\omega _{p})$ is computed for
various temperatures from an average over atomic positions within the well.}
\label{fig4}
\end{figure}

In support of these assertions, Fig.~\ref{fig4}(a) explores the
theoretical dependence of the transmission spectrum $T_{1}(\omega
_{p})$ on the set of FORT\ wells
selected, and hence on the distribution of values for $|\psi (\mathbf{r}%
_{FORT})|$ in the ideal case $\mathcal{T}=0$. Extending the average beyond
the $30$ \textquotedblleft best\textquotedblright\ FORT wells (i.e., $|g(%
\mathbf{r}_{FORT})|/g_{0}\lesssim 0.87$) leads to spectra that are
inconsistent with our observations in
Figs.~\ref{fig2},~\ref{fig3}. Figure~\ref{fig4}(b) likewise
investigates the dependence of $\bar{T}_{1}(\omega _{p})$ on the
temperature $\mathcal{T}$ for an atom trapped in the
\textquotedblleft best\textquotedblright\ well (i.e.,  $|\psi (\mathbf{r}%
_{FORT})|=1$ ). For temperatures $\mathcal{T}\gtrsim
200~\mathrm{\mu K}$, the calculated spectra are at variance with
the data in Figs.~\ref{fig2},~\ref{fig3}, from which we infer
atomic localization $\Delta z\simeq 33~\mathrm{nm}$\ in the axial
direction and $\Delta x=\Delta y\simeq 3.9~\mathrm{\mu m}$ in the
plane transverse to the cavity axis. Beyond these conclusions
drawn from Figs.~\ref{fig2}--\ref{fig4}, a consistent feature of
all our measurements is that reasonable correspondence between
theory and experiment is only obtained by restricting
$|\psi(\mathbf{r})|\gtrsim 0.8$.

Our experiment represents an important advance in the quest to
obtain single atoms trapped with optimal strong coupling to a
single mode of the electromagnetic field. The vacuum-Rabi
splitting is the hallmark of strong coupling for single atoms and
photons, and all measurements until now have required averaging
over many atoms for its observation. By contrast, we are able to
observe spectra $T_{1}(\omega _{p})$ on an atom-by-atom basis with
clearly resolved normal-mode splittings. These spectra contain
detailed quantitative information about the coherent coupling
$g(\mathbf{r})$ and FORT shifts for each atom. This information
indicates that the coupling $g$ is in a narrow range of
near-maximal values. Our observations are made possible by the
implementation of a new scheme to cool both the radial and axial
atomic motion. The capabilities demonstrated in this Letter should
provide the tools necessary to implement diverse protocols in
quantum information science \cite
{pellizzari95,duan04,cirac97,briegel00,mabuchi02}.

We gratefully acknowledge the contributions of T. Northup. This
research is supported by the Caltech MURI Center for Quantum
Networks, by the National Science Foundation, and by the Advanced
Research and Development Activity (ARDA).


\begin{thebibliography}{99}
\bibitem{birefringence} Stress-induced birefringence in the
cavity mirrors \cite{hood01} leads to two nondegenerate cavity modes
with orthogonal polarizations $\hat{l}_{\pm }$ and with mode
splitting $\Delta
\nu _{C_{1}}=4.4\pm0.2~\mathrm{MHz.}$ $\mathcal{E}_{FORT},$ $\mathcal{E}%
_{Raman}$ are linearly polarized and aligned along $\hat{l}_{+}$\
for the higher frequency mode.
\bibitem{T-define} $T_{1}(\omega _{p})$ is proportional to the ratio of photon flux
transmitted by $M_{2}$ to the flux $|\mathcal{E}_{p}|^{2}$ incident upon $%
M_{1}$ and is normalized such that the transmission $T_{0}(\omega
_{C_{1}})\equiv 1$\ for the empty cavity at line center.
Figs.~\ref{fig2},~\ref{fig3} also show $\langle n(\omega_p)\rangle$,
the corresponding intracavity photon number.
\bibitem{probepol} Relative to the polarization eigen-axes $\hat{l}_{\pm }$\
of the two cavity modes, the linear polarization vector
$\hat{l}_{p}$ for the probe field $\mathcal{E}_{p}$ is aligned along
a direction $\hat{l} _{p}=\cos \theta \hat{l}_{+}+\sin \theta
\hat{l}_{-}$, where $\theta =13^{\circ }$ for Fig.~\ref{fig2};
however, both theory and experiment are relatively insensitive to
$\theta $\ for $\theta \lesssim 15^{\circ }$.
\bibitem{efficiency} The efficiency for photon escape from the cavity is $\alpha _{e2}=0.6\pm
0.1$. The propagation efficiency from $M_{2}$ to detectors
$(D_{1},D_{2})$ is $\alpha _{P}=0.41\pm .03$, with then each
detector receiving half of the photons. $(D_{1},D_{2})$ have quantum
efficiencies $\alpha _{P}=0.49\pm 0.05$.
\bibitem{asymmetry} The maximum ac-Stark shifts of the $(F^{\prime}=5^{\prime}, m^{\prime}_F)$ states are given by $\{m_{F}^{\prime },U_{m_{F}^{\prime
}}\}=$ $\{\pm 5,1.18U_{0}\}$, $\{\pm 4,1.06U_{0}\}$, $\{\pm
3,0.97U_{0}\}$, $\{\pm 2,0.90U_{0}\}$, $\{\pm 1,0.86U_{0}\}$,
$\{0,0.85U_{0}\}$. These Stark shifts in conjunction with optical
pumping caused by $\mathcal{E}_p$ lead to the asymmetric peak
heights observed in Figs.~\ref{fig2},~\ref{fig3} via an effective
population-dependent shift of the atomic resonance frequency.
\bibitem{selection} Transmission spectra $T_{1}(\omega _{p}),\bar{T}%
_{1}(\omega _{p})$ are insensitive over a range of selection criteria $%
0.02\leq T_{thresh}\leq 0.73$.
\bibitem{nonthermal} Although the spectra are consistent with a
thermal distribution, we do not exclude a more complex model
involving probe-dependent heating and cooling effects.
\end{thebibliography}
\end{document}